# Multi-color broadband visible light source via three-dimensional GaN hexagonal annular microstructures

Young-Ho Ko[1,+], Jie Song[2], Benjamin Leung[2], Jung Han[2], and Yong-Hoon Cho[1,*]

[1]*Department of Physics and KI for the NanoCentury, Korea Advanced Institute of Science and Technology (KAIST), Daejeon 305-701, Republic of Korea*
[2]*Department of Electrical Engineering, Yale University, New Haven, Connecticut 06520, USA*
(Dated: February 24, 2014)

Broadband visible light emitting, three-dimensional hexagonal annular microstructures with InGaN/GaN multiple quantum wells (MQWs) are fabricated via selective-area epitaxial growth. The single hexagonal annular structure is composed of not only polar facet of (0001) on top surface but also semi-polar facets of {10$\bar{1}$1} and {11$\bar{2}$2} in inner and outer sidewalls, exhibiting multi-color visible light emission from InGaN/GaN MQWs formed on the different facets. The InGaN MQWs on (0001) facet emits the longer wavelength (green color) due to the larger well thickness and the higher In composition, while those on semi-polar facets of {10$\bar{1}$1} and {11$\bar{2}$2} had highly-efficient shorter wavelength (violet to blue color) emission caused by smaller well thickness and smaller In composition. By combining the multiple color emission depending on different facets, high efficiency broadband visible light emission could be achieved. The emission color can be changed with excitation power density owing to the built-in electric field on the (0001) facet, which is confirmed by time-resolved luminescence experiments. The hexagonal annular structures can be a critical building block for highly efficient broadband visible light emitting sources, providing a solution to previous problems related to the fabrication issues for phosphor-free white light emitting devices.

Keywords: InGaN, broadband visible light, multi-facet, selectively-area growth, light-emitting diode

---

[+] Current Address: Electronics and Telecommunications Research Institute (ETRI), Daejeon 305-700, Republic of Korea

[*] Email address: yhc@kaist.ac.kr

# 1. Introduction

The group III-nitrides, as representative materials for light-emitting diodes (LEDs), has attracted a wide range of attention in science and technology.[1-3] The broadband emission and multi-color emission have been important issue in the solid-state lighting.[4-6] The conventional white LEDs generally combined with phosphors had unavoidable energy conversion loss and poor color-rendering index. So, there have been various approaches to obtain broadband emission of InGaN without using phosphors.[7-9] The GaN micro-structures such as micro-rods, micro-stripes and micro-pyramids could be considered as important building blocks for realizing broadband and multi-color emission. It was possible to obtain multi-color emission from InGaN/GaN multi-quantum wells (MQWs) on the GaN multi-facets of the micro-structures, which were grown by the selective-area epitaxial growth (SAG) technique.[10-13]

T. Wunderer *et al.* studied three-dimensional structures of GaN providing multi-facets to obtain broadband emission. The InGaN/GaN MQWs on the triangular stripe structures of GaN were demonstrated as electrically-driven LEDs. The specific geometry influenced variation of thickness and In-composition, resulting in broadband emission.[14] To obtain multi-color emission with the triangular stripe structures, there were many approaches by combining different structures of the planar and the triangular structures, which had different wavelength emissions.[15,16] But there were difficulties to fabricate electrodes on the micro-structures because of the different height of the planar and the stripe structures. GaN nano-rod structures could be applied as the LED of the multi-color emission. Y. Hong *et al.* fabricated nano-rod based LED by growing InGaN/GaN MQW on the GaN nano-rod which had multi-color emission from the multi-facets of nano-rod. Because the current path was varied along the height of the nano-rods, the emission color was varied with injection current density.[17] The nano-rod or micro-rod structures not only had the non-uniform current path, but also had the leakage current, which were severe problems for the electrical fabrication due to geometrically difficult fabrication process.

The GaN pyramid structures composed of semi-polar facets have been applied to optical devices. It has been widely studied because the pyramid structure provides not only a reduced strain and dislocation densities but also diminished polarization field, resulting in enhancement of recombination efficiency.[18-21] Recently, we have studied the optical properties of pyramid structures and its application of electrical demonstration. It provided good optical performance due to high quantum efficiency of quantum dot and semi-polar

MQWs in the GaN pyramids.[22] The broadband emission could be realized due to broad distribution of In-composition and well thickness in the pyramids. Therefore, it was possible to obtain white-color emission by combining planar and pyramid structures with optimizing the ratio of two structures.[23] It still had difficulties on the electric fabrication because of non-equivalent height between planar and pyramid structures.

To overcome the aforementioned problems, we proposed the hexagonal annular structure of GaN as a single micro-structure with equivalent height, which can have the multi-facet in both inner and outer sidewalls for broadband and multi-color emission. The hexagonal annular structures have been widely adopted for the investigation of the growth mode because the GaN hexagonal annular structure contains various polar- and semi-polar-facets.[24,25] From the multi-facet of hexagonal annular structure, we expected that the InGaN/GaN MQWs grown on the hexagonal annular structure provided broadband and multi-color emission. Moreover, it enabled us to investigate the optical characteristic of the semi-polar facets because the InGaN/GaN MQWs grow simultaneously on the $\{10\bar{1}1\}$ and $\{11\bar{2}2\}$ semi-polar facets of the hexagonal annular structures. The InGaN on the semi-polar facets have been widely studied for increasing the emission efficiency due to reduced electric polarization, resulting in an improvement of the electron and hole wave-function overlap. In this study, we successfully fabricated InGaN MQWs on GaN hexagonal annular structures of multi-facets to realize broadband and multi-color emission via SAG with ring patterned. We systematically carried out the comparative study of the structural and optical properties of InGaN on the semi-polar facets of $\{10\bar{1}1\}$ and $\{11\bar{2}2\}$.

## 2. Result and Discussion
### 2.1. Selective-Area Epitaxial Growth

A schematic illustration of the hexagonal annular structure is shown in **Figure 1a**. The hexagonal annular structure of GaN was grown by SAG with a ring-patterned mask on a 2 µm-thick GaN on a *c*-sapphire substrate by metal-organic chemical vapor deposition (MOCVD). The inside and outside of diameters for ring pattern were 6 µm and 12 µm, respectively. The center-to-center distance of ring pattern was 16 µm. The 10-period InGaN/GaN MQW layers were grown on the GaN hexagonal annular structures. We confirmed that the hexagonal annular structure was grown well with the uniform array and the clear surfaces from top-view scanning electron microscope (SEM) image as shown in Figure 1b. The bottom distance between face-to-face sidewalls of a hexagonal annular was

12.3 µm for outside facets and 3.1 µm for inside facets, and the height was ~ 2.8 µm. The magnified SEM image of a single hexagonal annular was shown in Figure 1c. It contained 6 sidewalls for outside and 12 sidewalls for inside, together with top plane. We determined that the crystal orientation of the sidewalls of the hexagonal annular were {10$\bar{1}$1} for outside sidewalls and {10$\bar{1}$1} and {11$\bar{2}$2} for inside sidewalls, and {0001} for top plane. Although the {10$\bar{1}$1} and {11$\bar{2}$2} are the representative semi-polar facets of GaN with similar inclined angle of 62° and 58°, there is a lack of study to characterize the comparative properties for two semi-polar facets. Because the InGaN on the semi-polar and polar facets were grown simultaneously in the hexagonal annular structure, it was possible to obtain multi-color emission and possible to carry out the comparative study for characteristic of semi-polar facets between {10$\bar{1}$1} and {11$\bar{2}$2}.

## 2.2. Spatially-Resolved Optical Properties

The cathodoluminescence (CL) of hexagonal annular structure was measured to characterize spatially-resolved optical properties. **Figure 2a** shows a bird's eye-view SEM image of a hexagonal annular structure where the white points indicated the exciation position for CL spectra. We obtained CL spectra at 300 K for semi-polar facets of inside and outside sidewall and top plane as shown in Figure 2b. The CL peak wavelength was measured as 412 nm for {11$\bar{2}$2} of inside sidewall, 440 nm for {10$\bar{1}$1} of inside sidewall, 425 nm for {10$\bar{1}$1} of outside sidewall, and 525 nm for (0001). We confirmed the emission region of each wavelength from monochromatic CL mapping images as shown in Figure 2c-e. Each facet shows different emission peak wavelengths within a wide range of visible light even though grown at the same time. Especially, the (0001) facet have longest wavelength of green-color emission. For inside two semi-polar facets, {10$\bar{1}$1} have longer wavelength than {11$\bar{2}$2}. The origin of different emission wavelength could be expected that each facet had different In-incorporation efficiency, different growth rate of InGaN (and the consequent different well thickness), and different strain status (and hence different built-in electric field). For the same facet of {10$\bar{1}$1}, the inside facet had longer wavelength than outside facet, which could be due to the geometric origin of differences of In-incorporation efficiency, well thickness, and built-in electric field. As a result, the InGaN on the hexagonal annular structure provided broadband and multi-color emission in visible wavelength from the mixture of the polar and semi-polar facets.

## 2.3. Structural Characteristics

To confirm the origin of each facet emitting different wavelength emission, we measured the well thickness and In-composition from the transmission electron microscope (TEM). **Figure 3**. shows the high-angle annular dark-field (HAADF) scanning TEM images for the MQWs of each facet. The well thickness was measured as 1.55 (± 0.20) nm for {10$\bar{1}$1}, 1.37 (± 0.12) nm for {11$\bar{2}$2} and 5.38 (± 0.15) nm for (0001). For the comparison of the semi-polar facets of inside sidewalls, {10$\bar{1}$1} had slightly larger well thickness than {11$\bar{2}$2}, resulting in red shifted emission for {10$\bar{1}$1}. The peak shift due to the variation of well thickness for our sample was estimated as around 2.5 nm.[26,27] But the experimental result of CL peak shift of 28 nm was much larger than calculated value. To confirm the effect of strain-induced piezoelectric field, we considered the peak position of GaN from the CL measured at 80 K. It was known that the near-band edge emission of GaN is shifted with the strain.[28] We obtained the CL spectra along the height for each facet. Because the CL peak position of GaN for {10$\bar{1}$1} and {11$\bar{2}$2} facets were almost same, we expected that the origin of CL peak shift was not due to the effect of bulit-in electric field. (cf. Supporting Information for a detailed discussion). Therefore, we concluded that the origin of different emission of InGaN on {10$\bar{1}$1} and {11$\bar{2}$2} facets was mostly difference of In composition. We carried out the energy dispersive spectrometer (EDS) to measure the In-composition. Because of the limitation of resolution for detecting a single layer of InGaN, we measureed the relative In contents for the same area of MQWs (dotted-line rectangles in Figure 3). InGaN on {0001} had the highest In content of 5.5 (± 1.4) % and InGaN on {10$\bar{1}$1} had higher In content of 3.0 (± 0.1) % than InGaN on {11$\bar{2}$2} of 2.7 (± 0.1) %. To verify the result, we grew the InGaN MQW on the triagular stripe structure of GaN with horizontal line pattern for {10$\bar{1}$1} facet and vertical line pattern for {11$\bar{2}$2} facet. (Supporting information Figure S2.) Because the CL spectra of InGaN on the horizontal stripes with {10$\bar{1}$1} had longer wavelength emission than the vertical stripes with {11$\bar{2}$2}, we concluded that {10$\bar{1}$1} facets have higher In incorporation efficiency than {11$\bar{2}$2}. The origin of peak shift between inside and outside {10$\bar{1}$1} facets might be due to difference of strain, which was discussed in Supporting Information. From the same analysis for the (0001) facet, we found that the InGaN MQWs on (0001) had higher In composition compared to the semi-polar facets. As a result, the hexagonal annular structure provided the multi-color emission orirginated from the multi-facets of polar and semi-polar facets.

## 2.4. Multi-color Broadband Visible Light Emission

The photoluminescence (PL) of a hexagonal annular structure was measured with changing the excitation power density at 10 K to characterize the optical properties, as shown in **Figure 4a**. There were two main emissions of InGaN on the semi-polar facets (401.9 nm at 25 mW) and on the polar facet (517.7 nm at 25 mW). The full width at half maximum (FWHM) of InGaN on semi-polar facets was ~ 226.8 meV and InGaN on polar facet was ~ 213.3 meV. We supposed that the broadband spectra with large FWHM originated from the mixture of multi-facets. The peak positions of InGaN on polar facet revealed blue-shift of 21.6 nm as increasing the excitation power density from 30 µW to 25 mW. Interestingly, however, the InGaN on semi-polar facets showed no shift with the excitation power density, which is direct indication of dramatically reduced internal electric field in the MQWs formed on semi-polar facets of hexagonal annular structure. Therefore, the emission-color of PL spectra were changed from (0.22, 0.45) to (0.34, 0.56) in the international commission on illumination (CIE) map, as shown in the inset of Figure 4a. This result provides the possibility to realize the color-tunable optical device with broadband emission. To investigate the transient carrier dynamics, we performed the time-resolved PL experiment with a femto-second Ti:Sapphire laser excitation of 340-nm wavelength. Figure 4b shows the decay lifetime of InGaN and a streak image. The decay lifetime of InGaN MQWs grown on semi-polar facets and polar facet were obtained from the streak image with the wavelength window of 380 ~ 440 nm and 510 ~ 580 nm, respectively. The decay lifetimes of InGaN MQWs on semi-polar facets and InGaN MQWs on polar facet were measured to be 168 ps and 1.83 µs, respectively. The decay lifetime at 10 K was considered as the radiative lifetime by assuming that 10 K was low enough to disregard the non-radiative processes. The decay lifetime of InGaN MQWs on semi-polar facets had much shorter than InGaN MQWs on polar facet, which was due to the smaller built-in electric field and thiner layers of InGaN on semi-polar facets resulting in higher recombination rate. Consequently, we realized the broadband and multi-color visible emission from the multi-facets of hexagonal annular, which had the advantages of high quantum efficiency of semi-polar facets. It was so meaningful because the hexagonal annular structure can provide the possibility for fabrication of optical devices as a single micro-structure avoiding many drawbacks of combined structures for the broadband and white-color emission.

## 3. Conclusion

We have successfully obtained the broadband and multi-color emission by growing InGaN/GaN MQWs on a GaN hexagonal annular structure. The (0001) polar facet as well as the {10$\bar{1}$1} and {11$\bar{2}$2} semi-polar facets were obtained simultaneously through the hexagonal annular structure grown on the ring patterned mask. From the hexagonal annular structure, we could investigate comparative study of polar and semi-polar facets with different well thickness and In incorporation efficiency. The InGaN on (0001) facet emitted the longest wavelength of 525 nm due to the largest well thickness and the highest In composition. The InGaN on {10$\bar{1}$1} facet had longer wavelength emission than the InGaN on {11$\bar{2}$2} facet because the {10$\bar{1}$1} facet had higher efficiency of In-incorporation than {11$\bar{2}$2}. The hexagonal annular structure emitted the broadband spectra with two main emissions originated from InGaN on polar and semi-polar facets. The emission color could be changed with excitation power density due to the built-in electric field on the (0001) facet. We confirmed that the InGaN MQWs formed on semi-polar facet had short decay lifetime of 168 ps, providing the high internal quantum efficiency. Furthermore, we expected to obtain the electrically-driven white LEDs by optimizing the well thickness and In composition. Therefore, the hexagonal annular structures will provide a solution to previous problems related to the fabrication of single structure emitting broadband and multi-color emission and it will be building blocks for highly efficient broadband visible light emitting devices.

**4. Experimental Section**

A hexagonal annular structure of GaN was grown by MOCVD through the SAG technique with a ring patterned mask. A 100 nm thick $SiO_2$ layer was deposited on a 2 µm thick *c*-GaN template grown on sapphire substrate. We obtained the ring patterned mask by photo lithography with the inside and outside diameter of 6 µm and 12 µm, respectively. The 10 periods InGaN/GaN MQW was grown at 750 °C on the GaN hexagonal annular structure. To observe the morphology of hexagonal annular structure, we employed the SEM (Hitachi S-4800) with a 10 kV acceleration voltage. We determined the crystal orientation of the outside sidewalls of the pyramids to be {10$\bar{1}$1} through the prediction of the crystal direction based on the primary flat zone of the sapphire substrate. We performed CL experiments (Gatan, Mono CL4) combined with SEM (FEI, XL 30S FEG), where the acceleration voltage was 5 kV to characterize the spatially-resolved optical properties. The CL spectra at 80 K were measured by using the cryogenic stage with the liquid nitrogen. For the TEM measurement, the cross section of MQWs on {10$\bar{1}$1} and {11$\bar{2}$2} facets in the hexagonal annular structures were prepared by using focused ion beam. The Cs-corrected TEM (JEOL, JEM ARM200F) was employed to obtain high resolution HAADF images. The time-integrated PL spectra were measured with a continuous-wavelength He-Cd laser of 325 nm at 10 K. To carry out the time-resolved PL experiment, a mode-locked Ti:sapphire laser (Coherent, Cameleon Ultra II) was used with doubling frequency. The wavelength, width, and repetition rate of the pulse laser was 340 nm, 200 fs, and 200 kHz, respectively. A streak camera (Hamamatsu, C7700-01) was employed to measure the decay lifetime. For the time-integrated PL and the time-resolved PL experiment, we used a cryogenic system, whereby the temperature of the samples was controlled from 300 K to 10 K.

**Acknowledgements**
This work was supported by the National Research Foundation (NRF-2013R1A2A1A01016914, NRF-2013R1A1A2011750) of the Ministry of Education, the Industrial Strategic Technology Development Program (10041878) of the Ministry of Knowledge Economy, and KAIST EEWS Initiative.

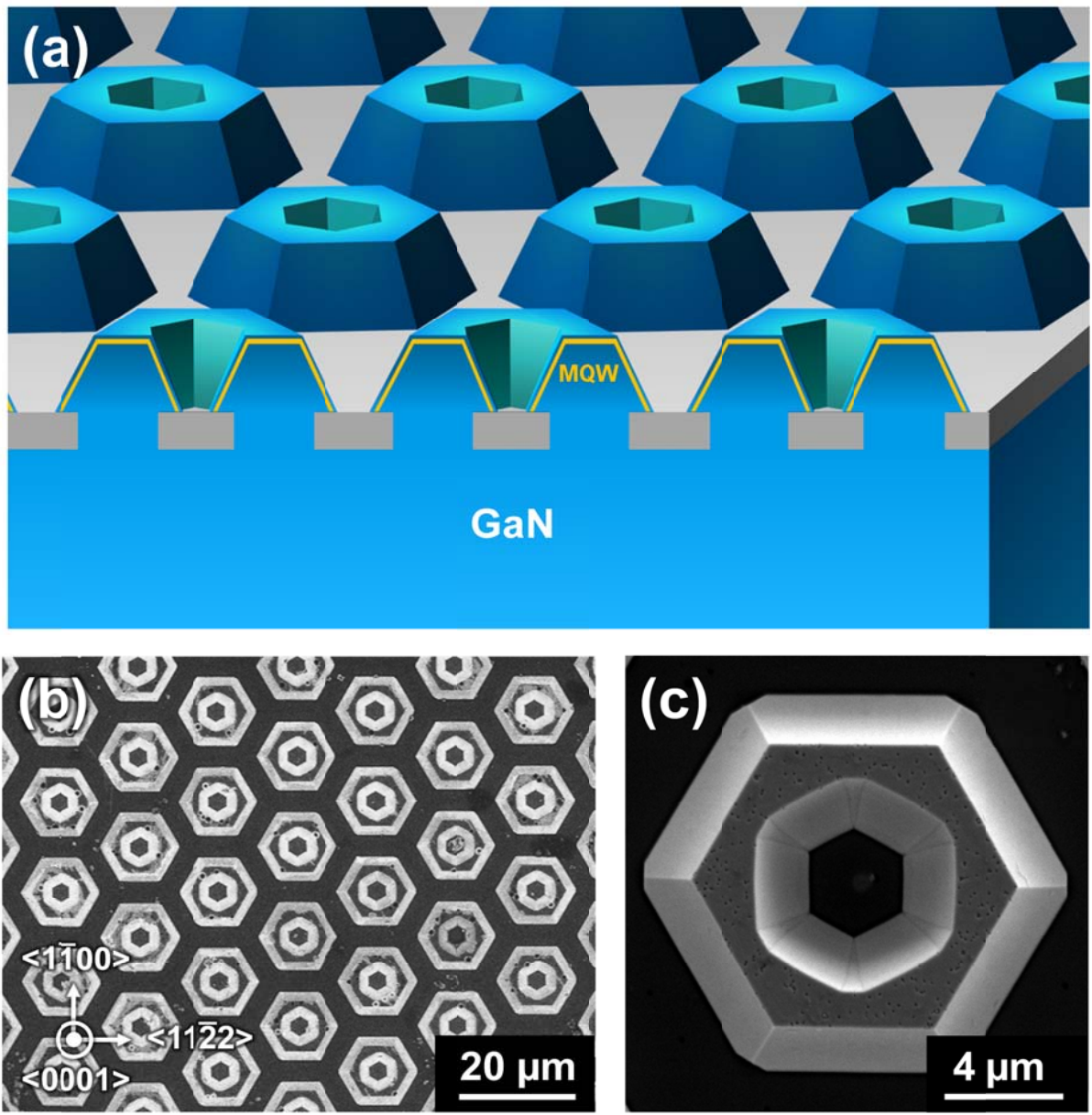

**Figure 1.** Schematic illustration of a) the hexagonal annular structure with InGaN/GaN MQWs. b) Top-view SEM image of the hexagonal annular structure, and c) the magnified image of single structure.

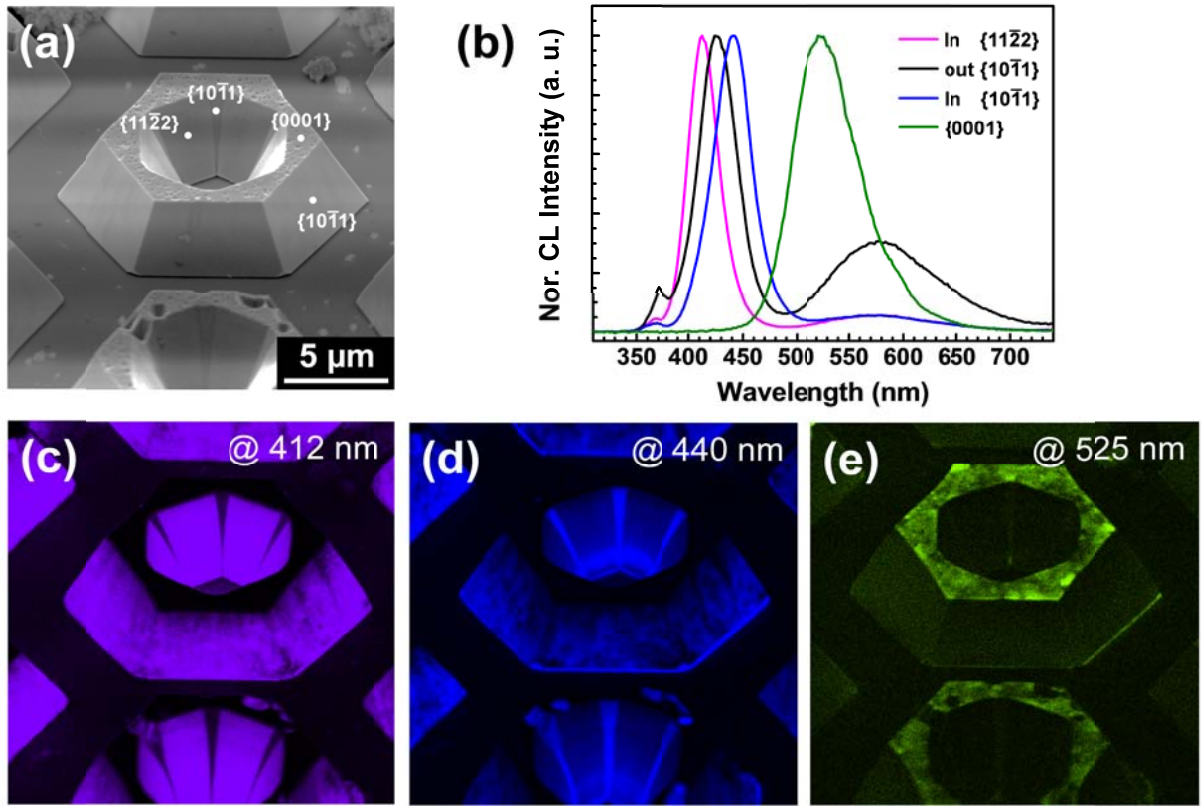

**Figure 2.** Structural and spatially-resolved optical characterization. a) Bird's eye view SEM image of hexagonal annular structure. The white points indicated the excitation positions of CL. b) The CL spectra of InGaN/GaN MQWs on the inside {11$\bar{2}$2}, outside {10$\bar{1}$1}, inside {10$\bar{1}$1} and {0001} facets. A monochromatic CL mapping image at a wavelength of c) 412 nm, d) 440 nm and e) 525 nm.

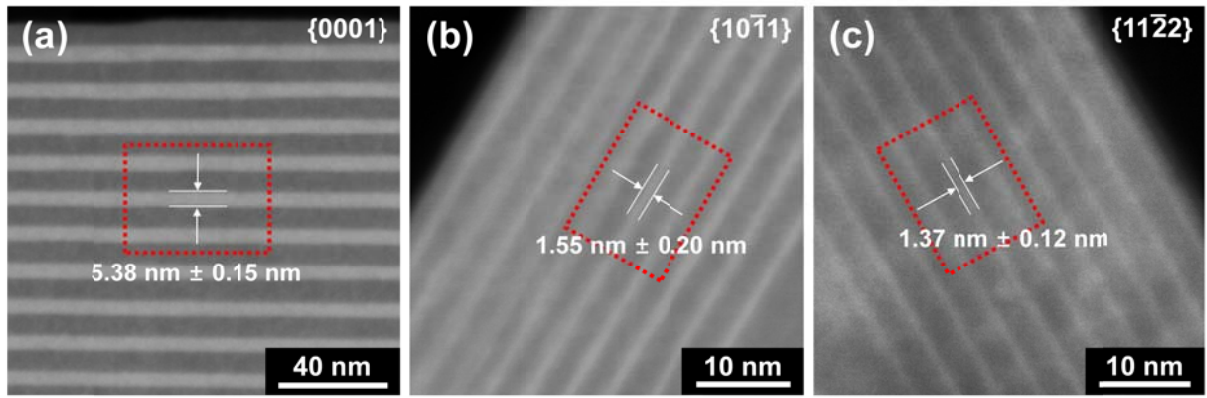

**Figure 3.** The TEM measurements of the InGaN MQWs on each facet. The cross-sectional HAADF STEM images of InGaN/GaN MQWs a) on the {0001}, b) on the {10$\bar{1}$1} and c) on the {11$\bar{2}$2} facets. The bright contrast indicated the InGaN layer.

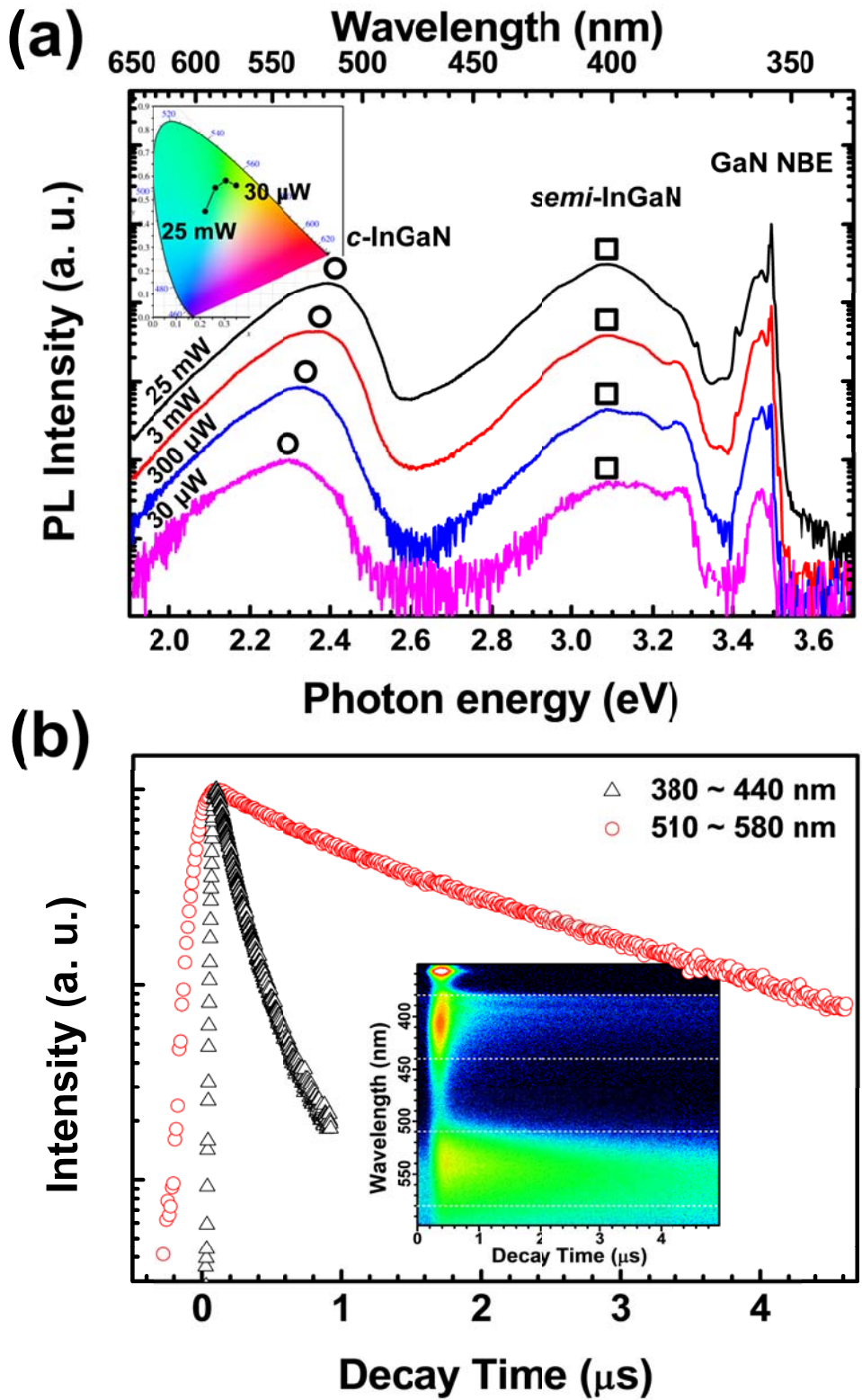

**Figure 4.** Time-integrated and time-resolved optical properties of the hexagonal annular structure. a) The PL spectra with changing the power density of the excitation laser. The insets show CIE map of each spectrum. b) The decay lifetime of MQWs on polar facet and semi-polar facets with the detection wavelength of 380 ~ 440 nm and 510 ~ 580 nm, respectively, which were displayed in the steak image measured at 10 K (inset).